\documentclass[prb,aps,twocolumn,showpacs,floats]{revtex4}

\usepackage{graphicx}
\usepackage{bm}
\usepackage{hyperref}
\usepackage{times}
\usepackage{amsmath2000}

\newcommand{\eq}[1]{Eq. (\ref{#1})}
\newcommand{\dd}{{\rm d}}
\newcommand{\denom}[1]{\left( r + \varpi^2 + q^2 + fq_{#1}^2 \right)}
\allowdisplaybreaks[2]

\begin{document}

\title{Theory of the Quantum Paraelectric -- Ferroelectric Transition}
\author{R. Roussev$^{(1)}$ and A. J. Millis$^{(2)}$}
\address{$^{(1)}$Center for Materials Theory, Department of Physics and
Astronomy, Rutgers University, 136 Frelinghuysen Rd, Piscataway, NJ 08854\\
$^{(2)}$Department of Physics, Columbia University, 538 W 120$^{th}$ St,
New York, NY 10027}

\begin{abstract}
A realistic theory of the quantum paraelectric -- ferroelectric transition
is presented, involving parameters determined from band calculations and a
renormalization group treatment of critical fluctuations.  The effects of
reduced dimensionality and deviations from cubic symmetry are determined.
Expressions for the pressure dependence of $T_c$ as well as $p$ and $T$
dependence of the specific heat are derived, and evaluated for realistic
materials parameters for the systems BaTiO$_3$ and PbTiO$_3$.  In these
materials the ferroelectric soft mode dispersion apparently exhibits
a very strong cubic anisotropy, which affects results in an important,
albeit quantitative, manner.  A change in order parameter orientation from
(100) to (111) is predicted as quantum criticality is approached.
\end{abstract}

\pacs{80.Bh, 77.84.Dy, 05.70.Jk, 64.60.-i}

\maketitle

\section{Introduction}
Ferroelectrics and the closely related high dielectric constant materials
are important in many areas of modern technology including memory,
sensor and electronic applications, and are of fundamental scientific
interest \cite{Lines77}.  The ferroelectric phase change belongs to a
class of structural transitions, generally termed {\em ferrodistortive},
triggered by zone-center soft modes of lattice motion.  Characteristically
the ferroelectric transition involves the condensation of an optically
active lattice mode which causes the appearance of long-range polar order
and the breaking of the inversion symmetry of the `high-temperature'
prototype lattice.  One important issue is the ferroelectric quantum
critical point, {\em i.e.} the physics occurring when, by varying a control
parameter $r$ (applied pressure or change of chemical composition), the
transition temperature of a ferroelectric is driven to zero.  Although a
quantum critical point occurs at $T=0$, the fluctuations associated with the
critical point may control behavior over a range of temperature and pressure.
For example K. A. M\"uller \cite{Mueller79} coined the term ``quantum
paraelectric'' to describe materials in which ferroelectric ordering is
prevented by quantal fluctuations.  A material just on the disordered
side of the $T=0$ ferro-paraelectric transition is therefore an example
of a ``quantum paraelectric''.  The important feature of quantal ($T=0$)
phase transitions is that temporal fluctuations must be treated on the same
footing as thermal ones \cite{Hertz73,Sachdev99}.  This raises the effective
dimensionality and makes the critical behavior more mean-field-like but
with temperature dependence controlled by `dangerous irrelevant operators'
\cite{Millis93}.

Ferroelectric transitions may be described by bosonic field theories
with undamped dynamics (if the effect of free carriers may be neglected)
and complicated dispersions arising from the long range of the dipolar
interaction.  Quantum critical phenomena associated with undamped bosonic
field theories with short ranged interactions have been extensively
studied \cite{Sachdev99}.  The effect of long range (dipole) forces
was studied by Rechester \cite{Rechester71} and by Khmel'nitskii and
Shneerson \cite{Khmelnitskii71} within an approximation equivalent to the
self-consistent one loop approximation of Moriya \cite{Moriya85}.  Aharony
and Fisher \cite{Fisher73} studied the classical ferroelectric transition
and found that anisotropies associated with the dipolar interaction led
to a new universality class.  In this paper we reexamine the issue in
light of recent developments in the theory of quantum critical phenomena.
We formulate a realistic action for the ferroelectric soft modes, show how
estimates of the parameters may be obtained from ab initio calculations and
study quantitatively the consequences of the dipolar-induced anisotropies.
Our results agree in essentials with those of Rechester and Khmel'nitskii
and Shneerson, but we obtain a more detailed and quantitative picture of
the phase boundary, of the effect of anisotropy, and of the logarithmic
corrections arising at the marginal dimensionality, which lead to an evolution
of the anisotropy as the ordered phase is approached.

\section{Order parameter and action}
The order parameter of our theory is the local polarization ${\bm \phi}$.
Taking into account the effective dipole charges $e_i^*$ of the soft modes,
${\bm \phi}$ can be formally written
\begin{equation}
{\bm \phi}(x, t) = \sum_{i=1}^5 e_i^* {\bf r}_i(x, t). \label{ordparam}
\end{equation}
Here the index $i$ runs through the atoms of the unit cell of the prototype
perovskite lattice with stoichiometry ABO$_3$; ${\bf r}_i$ are the vector
displacements of each atom.

We now write a Ginsburg -- Landau action describing quantal and thermal
fluctuations of ${\bm \phi}(x, t)$. The crucial point is that because ${\bm
\phi}$ corresponds to a dipole fluctuation it generates electric fields
which lead to a long-range interaction. We have (in space and imaginary time)
\begin{eqnarray}
S[\phi_\alpha(x, \tau)] & = & \int {d^d x \over a^d} \int_0^{\hbar /T}
d\tau E_0 \Big[ {a^2 \over c^2} \left( \partial_\tau \phi_\alpha(x, \tau)
\right)^2 \nonumber \\
  & + & a^2 \left( \nabla \phi_\alpha(x, \tau) \right)^2 + \phi_\alpha(x,
\tau) r_{\alpha \beta} \phi_\beta(x, \tau) \nonumber \\
  & + & \int {d^d x' \over a^d} \phi_\alpha(x, \tau) F_{\alpha\beta}(x -
x') \phi_\beta(x', \tau) \label{action} \\
  & + & \sum_{\alpha\beta} \left( u + v_\alpha \delta_{\alpha\beta}
\right) \phi_\alpha^2(x, \tau) \phi_\beta^2(x, \tau) \Big] +
... \nonumber
\end{eqnarray}
Here $a$ is the lattice constant, $c$ is the speed of the phonons in the
softest direction, and $E_0 = \hbar c(\pi/a)$ is the typical energy scale
of ferroelectric fluctuations in the (100) direction; our choice of units
is such that the field, mass and coupling constants are dimensionless.
The term proportional to $F_{\alpha\beta} (x) = (d-2) (x^2 \delta_{\alpha
\beta} - dx_\alpha x_\beta)/x^{d+2}$ represents the dipole interaction. In
momentum space
\cite{Fisher73}
\begin{eqnarray}
F_{\alpha\beta} (q) & = & \int d^d x F_{\alpha\beta} (x) e^{i{q \over a}
\cdot x} = \left(r_{0\alpha} + f_\alpha q_\alpha^2 \right) \delta_{\alpha
\beta} \label{ewald} \\
  &  & + \left( g_{\alpha\beta} - q^2 h_{\alpha\beta} \right) {q_\alpha
q_\beta \over q^2} + {\mathcal O}(q^4), \nonumber
\end{eqnarray}
where $r_{0\alpha}$, $f_\alpha$, $g_{\alpha\beta}$, and $h_{\alpha\beta}$
depend on details of the underlying lattice.  We assume that the non-local
quadratic terms represented by $g_{\alpha\beta}$ and $h_{\alpha\beta}$
obey the same symmetry as the local quartic interaction terms $u+v_\alpha
\delta_{\alpha\beta}$.  Thus, in general, we will have $g_{\alpha\beta} =
g + g_I \delta_{\alpha\beta}$; $g_I > 0$ lowers the symmetry to Ising.
The term $r_{0\alpha}$ combined with {\em local} bare mass terms
makes up $r_{\alpha\beta}$ in \eq{action}.  We shall consider cubic and
tetragonal symmetry, so $r_{\alpha\beta} = r_\alpha \delta_{\alpha\beta}$.
In \eq{ewald} and in all of the following we use dimensionless momenta
$q_\alpha \in [-\pi, \pi]$.

Diagonalization of the quadratic part of the action yields the phonon modes,
and the paraelectric -- ferroelectric transition occurs when the lowest zone
center mode frequency vanishes.  The gradient term in \eq{action}, along with
$f_\alpha$ and $h$, controls the dispersion of modes.  Note that $f_\alpha
\neq 0$ implies an anisotropic derivative $\sum_\alpha(\nabla_\alpha
{\bm \phi}_\alpha)^2$; in a spherically symmetric system $f_\alpha=0$.
For simplicity we refer to the case $g_I = 0$, $f_\alpha > 0$ as Heisenberg
also, because the order parameter exhibits a continuous rotational symmetry.
Previous renormalization group studies of the classical paraelectric --
ferroelectric transition have treated the $f_\alpha$-terms as a small
perturbation \cite{Fisher73,Khmelnitskii71}.  Khmel'nitskii and Shneerson
\cite{Khmelnitskii71} argue that although $f_\alpha$ in typical materials
(e.g.  BaTiO$_3$) is of the same order of magnitude as $g_{\alpha\beta}$ and
$h_{\alpha\beta}$, the anisotropy of observable quantities is usually weak.
Because band theory calculations indicate that in many ferroelectric systems
$f_\alpha > 1$ are quite large we present here a treatment valid for any $f$.

The $u$ and $v_\alpha$ terms represent local anharmonic interactions.
The materials of main interest here have cubic symmetry in which case
$v_\alpha = v$.  The quartic interaction in \eq{action} (dropping momentum
and energy integrals for simplicity) becomes
\begin{equation} \label{quartic}
S^{(4)}[{\bm \phi}] = u \left( \sum_\alpha \phi_\alpha^2 \right)^2 +
v \sum_\alpha \phi_\alpha^4
\end{equation}
The term proportional to $u$ is rotationally invariant and insensitive
to the polarization orientation, and the sign of the {\em second} term
determines the polarization orientation in the ordered phase.  At the mean
field level the action \eq{action} is minimized by polarization of magnitude
\begin{equation}
P^2 = \sum_\beta \phi_\beta^2 = - {dr \over 2(du+v)}
\end{equation}
When $v<0$ \eq{action} is minimized by a polarization along (111) with
$\phi_x = \phi_y = \phi_z = P/d$ whereas for $v>0$ the polarization is along
(100) with $\phi_x = \phi_y = 0$, $\phi_z = P$.  The values of the quartic
interaction are in each case
\begin{equation}
S^{(4)}[{\bm \phi}] = {(dr)^2 \over 4(du+v)} \times
	\begin{cases}
	\left( u + v/27 \right), & v<0 \\
	\left( u + v \right), & v>0
	\end{cases}
\end{equation}
The condensation energy is of order $r^2/u$ and as $v \to 0$ the energy
barrier separating different symmetry-allowed polarization directions is a
factor of order $v/u$ smaller than the condensation energy.  The condition
for the stability of a quartic interaction is the positive definiteness
of \eq{quartic} which (in cubic symmetry, dimensionality $d$ and at the
mean field level) translates into
\begin{subequations} \label{initconditions}
\begin{align}
u + v &> 0 \\
d u + v &> 0 
\end{align}
\end{subequations}
If these conditions fail, sixth order terms in ${\bm \phi}$ have to be
included and the transition may be first-order.

\begin{figure}
\includegraphics[scale=0.85]{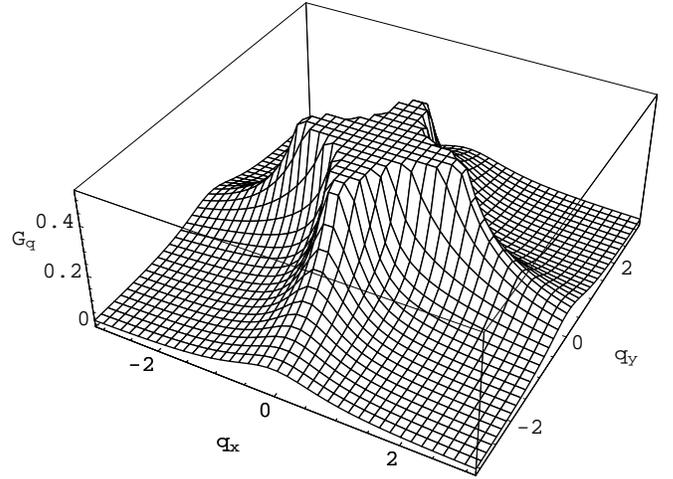}
\caption{Generic ferroelectric propagator \eq{propagator2d} in the static
limit $\varpi = 0$ with cubic symmetry and large anisotropy; $q_z = 0$,
$r=0$, $f=5.0$, $\Lambda = \pi$.} \label{propGraph}
\end{figure}

\section{Propagator and modes}
\eq{action} and \eq{ewald} define a model for the phase transition
in a ferroelectric near a quantum critical point.  In the absence of
nonlinearities, the Heisenberg order parameter correlation function
$G^0_{\alpha\beta} = \langle \phi_\alpha \phi_\beta \rangle_0$ is
\cite{Fisher73}
\begin{subequations}
\begin{align}
G_{\alpha \beta}^0 & = {1 \over r_\alpha + \varpi^2 + q^2 +
f_\alpha q_\alpha^2 } \label{propagator} \\
	&\times \left( \delta_{\alpha \beta}
- { (g -h q^2) q_\alpha q_\beta \over \left[ q^2 +
(g - h q^2)Q \right] [ r_\beta + \varpi^2 + q^2 + f_\beta
q_\beta^2 ]} \right) \nonumber \\
Q		   & = \sum_\gamma {q_\gamma^2 \over	r_\gamma + \varpi^2
+ q^2 + f_\gamma q_\gamma^2} \label{Q}
\end{align}
\end{subequations}
Here $\varpi = 2n \pi (T/E_0)$ is a dimensionless bosonic Matsubara
frequency, and in our conventions $G$ is dimensionless.

The nature of the modes defined by the poles of \eq{propagator} can be
best understood by considering the polarization of the ferroelectric
fluctuation vector ${\bm \phi}$.  For every ${\bf q}$ there are $d-1$
transverse and one longitudinal polarizations, all orthogonal to each other.
The longitudinal mode is always stiff with $\varpi_\parallel = {\mathcal O}
(g)$, and $h$ only enters the dispersion of the longitudinal mode; both
$g$ and $h$ are irrelevant to the critical behavior.  The remaining $d-1$
modes are soft and in the case of cubic symmetry have the general dispersion
\begin{equation}
\varpi^2_\lambda (q) = r_\lambda + q^2 \left[ 1 + f A_\lambda(\Omega_q)
\right] \label{generalModes}
\end{equation}
where $\Omega_q$ is the set of angles defining the direction of ${\bf q}$
and $A_\lambda$ are lengthy expressions derived from \eq{propagator}.
\eq{generalModes} includes all modes with the convention $r_\perp = r$,
$r_\parallel = r + g$.  For all ${\bf q}$ such that the polarization of
a transverse mode points along a crystal axis the respective dispersion
softens additionally ($A_\lambda(\Omega_q) = 0$) if $f>0$.  The effect of
$f$ is most easily seen by setting $q_z = 0$ and considering only the $XY$
block. The resulting transverse mode propagator
\begin{equation}
G(\varpi, q) = \left( r + \varpi^2 + q_x^2 + q_y^2 + 2f {q_x^2 q_y^2 \over
q_x^2 + q_y^2} \right)^{-1} \label{propagator2d}
\end{equation}
is shown in Fig.  (\ref{propGraph}).  The dipolar anisotropy ($f$) leads
to ridges suggestive of quasi one dimensional behavior.

\begin{table}[!hbp]
\begin{ruledtabular}
\begin{tabular}{lllccc}
& $\hbar c$ [$m$eV-\AA] & $E_0$ [$m$eV] & $f$ & $u$ & $v$ \\
\hline
BaTiO$_3$ &6.36 &5.00 & 4.7 & 1.25 & 0.68 \\
PbTiO$_3$ &6.79 &5.37 & 1.1 & 0.26 & -0.09
\end{tabular}
\end{ruledtabular}
\caption{Numerical values of the parameters used in the action \eq{action}.
The values for $u$ and $v$ are {\em initial conditions} for the RG flow of
the two interaction constants in \eq{action}.}
\label{parameters}
\end{table}

\section{Interaction renormalization}
\begin{figure}
\center{\includegraphics{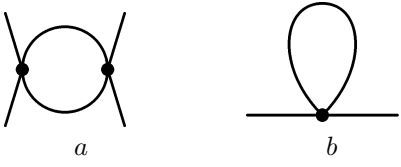}}
\caption{One loop diagrams for the renormalization of (a) the quartic
interaction $u$ and (b) the masses $r_\alpha$.}
\label{diagrams}
\end{figure}

We obtain the parameters $c$, $r_\alpha$, $f_\alpha$, $g$  and $h$ in
\eq{action} by fitting the poles of \eq{propagator} to first-principles
phonon dispersion curves such as those in Ref.  \cite{Ghosez99}.  We fit
the numerically calculated mode frequencies near the zone center to the
modes predicted by \eq{propagator} along crystal symmetry directions.
The soft modes' speed $c$ and mass $r(T=0)$ at the lattice constants used in
Ref. \cite{Ghosez99} (ambient pressure) are readily obtained by fitting the
dispersions along (100) to $\omega = c \sqrt{r/a^2} + (1/2)c\sqrt{r/a^2}
(q/a)^2 + {\mathcal O} (q^4)$.  The anisotropy parameter $f$ is obtained
from direct ratios of the curvature of the dispersion along (110) and (111).

The size of the interaction constants $u$ and $v$ can be estimated from
first-principles variational studies of a Landau free energy of the
system \cite{Vanderbilt94} $E({\bf w}) = \kappa {\bf w}^2 + \alpha'
{\bf w}^4 + \gamma' \sum_{\alpha > \beta} w_\alpha^2 w_\beta^2$ where
$E({\bf w})$ is the free energy per unit cell and ${\bf w}$ is a soft
mode lattice displacement.  The parameters $\alpha'$ and $\gamma'$
(Table V in \cite{Vanderbilt94}) studied by these authors are related
to $u$ and $v_\alpha = v$ in \eq{action} by $u = (\alpha' + \gamma'/2)
(r/\kappa)^2 E_0$ and $v = -(\gamma'/2) (r/\kappa)^2 E_0$.  For BaTiO$_3$
and PbTiO$_3$ (which have cubic symmetry so $v_\alpha = v$ and $f_\alpha =
f$) we find the results listed in Table \ref{parameters}.

\begin{figure}
\hspace{-4em} \includegraphics{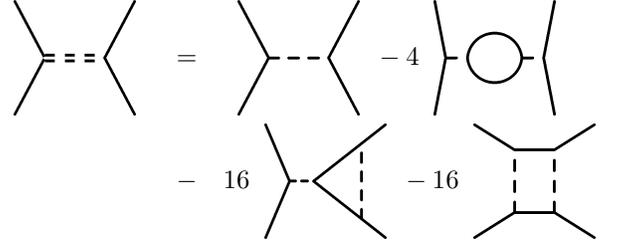} \label{vertex}
\caption{Renormalization diagrams to tree and one-loop order
for the anisotropic interaction parameter $u_{\alpha\beta} = u +
v\delta_{\alpha\beta}$ in \eq{action}.  This is the diagrammatic
representation of \eq{rgint}.}
\end{figure}

We now study the relevance of quartic interactions in the vicinity of
the critical point.  The one-loop correction to $u$ and $v$ is given
generically by the diagram in Fig. (\ref{diagrams}a) and the respective
renormalization equation \cite{Aharony73} is given by
\begin{multline} \label{rgint}
\sum_{\alpha\beta} u_{\alpha\beta}^{l+1} \phi^\alpha \phi^\alpha
\phi^\beta \phi^\beta = \zeta_l^4 b^{-3(d+z)} \Big[ \sum_{\alpha\beta}
u_{\alpha\beta}^l \phi^\alpha \phi^\alpha \phi^\beta \phi^\beta \\
- 4 \sum_{\alpha\beta\gamma\delta} u_{\alpha\gamma}^l u_{\beta\delta}^l
\big(w_{\gamma\delta\gamma\delta} \phi^\alpha \phi^\alpha \phi^\beta
\phi^\beta \\ 
+ 4w_{\gamma\beta\gamma\delta} \phi^\alpha \phi^\alpha \phi^\beta \phi^\delta
+ 4w_{\alpha\beta\gamma\delta} \phi^\alpha \phi^\beta \phi^\gamma \phi^\delta
\big) \Big]
\end{multline}
where $u_{\alpha\beta} = u + v\delta_{\alpha\beta}$, the external momentum
integrations are omitted for brevity, and $w_{\alpha\beta\gamma\delta}$
are the one-loop integrals over fast modes
\begin{equation} \label{wsdef}
w_{\alpha\beta\gamma\delta} = T\sum_{\varpi_n} \int {\dd^d q \over (2\pi)^d}
G^0_{\alpha\beta} (\varpi, q) G^0_{\gamma\delta}(\varpi, q).
\end{equation}
In \eq{rgint} $\zeta_l = b^{1+d/2}$ is the field renormalization when
the fast modes in the shell $\Lambda/b < q < \Lambda$ are integrated out.
The diagrammatic version of \eq{rgint} is shown in Fig. (\ref{vertex}).
In $d=3$ the system is in its marginal dimension and the prefactor in
\eq{rgint} is $\zeta_l^4 b^{-3(d+z)}$ so that the leading interaction
renormalization is quadratic.  The generic form of the renormalization
equations is
\begin{equation}
-\delta u \sim u^2 \int_{-\infty}^\infty {\dd \varpi \over 2\pi} \int^\pi
{\dd^d q \over (2\pi)^d} G^2(\varpi, q) \coth {\varpi \over 2T}, \label{uRG}
\end{equation}
where $G$ is a soft eigenmode of the Gaussian ferroelectric propagator.
We first consider the simplest case of isotropic interactions ($v_\alpha=0$)
in an isotropic medium ($f_\alpha = 0$, $r_\alpha = r$) in the low
temperature limit, and we also let $h=0$ for simplicity.  As explained
above, the correlation function \eq{propagator} then has a Heisenberg-like
rotationally invariant form with $d-1$ soft eigenmodes $G^{-1} = r +
\varpi^2 + q^2 = r + Q^2$ and one stiff (non-critical) eigenmode $G^{-1} =
r + Q^2 + g$.  Including only the soft eigenmodes in \eq{uRG} the recursion
relation for $u$ is respectively
\begin{equation}
- \delta u \sim \int^\pi {u^2 Q^d \over (r + Q^2)^2} dQ \sim u^2
	\left\{ \begin{array}{lc}
		  \log 1/r, & \quad d=3 \\
		  1/\sqrt r, & \quad d=2 \\
		  1/r,       & \quad d=1.
	        \end{array} \right. \label{heisU}
\end{equation}
\eq{heisU} shows that upper critical dimension of a Heisenberg isotropic
quantum critical ferroelectric system is $d_c = 3$.

For a uniaxial (Ising-like) ferroelectric there is a preferred `easy axis'
for the orientation of ${\bm \phi}$ which brings about a further increase
in effective dimensionality \cite{Fisher73}. The respective correlation
function has the form $G(\varpi, q) = (r+\varpi^2 + q^2+g_I q_z^2/q^2)^{-1}$
which gives for $\delta u$
\begin{eqnarray}
- \delta u & \sim & u^2 \int \dd \varpi \int \dd q \int{q^{d-1} \dd \cos
\theta \over (r + \varpi^2 + q^2 + g_I \cos^2 \theta)^2 } \label{uniaxU} \\
	& \sim & {u^2 \over g_I^{1/2}}
\int^\pi {Q^{d+1} \dd Q \over (r + Q^2)^2} \sim {u^2 \over g_I^{1/2}}
	\left\{ \begin{array}{lc}
		  r^{1/2}, & \qquad d=3 \\
		  \log 1/r, & \qquad d=2.
	        \end{array} \right. \nonumber 
\end{eqnarray}
Thus the upper critical dimension of an Ising isotropic quantum critical
ferroelectric system is reduced to $d_c = 2$.  In the case of a preferred
`easy plane' of polarization the propagator has the two eigenmodes $G = (r +
\varpi^2 + q^2)^{-1}$ and $G = (r + \varpi^2 + q^2 + g\sin^2 \theta)^{-1}$
so that the $XY$ -model and Heisenberg ferroelectrics have identical
coupling renormalizations which is due to the existence of the same soft
mode in both cases.

We now study the possibility that the quasi one dimensional behavior
associated with $f >> 1$ may modify the criticality.  We illustrate the
issues using the notationally simpler $d=2$, $g \to \infty$ case, and have
verified that our results hold in $d=3$ also.  From Fig. (\ref{diagrams}a)
and \eq{propagator2d} we obtain, after integration over $\varpi$ and the
magnitude of $q$
\begin{eqnarray}
- \delta u & = & \left( {u \over 4\pi} \right)^2 \int_0^{2\pi} {\dd \varphi
\over 1 + (f/2) \sin^2 2\varphi} \label{anisotropicU2d} \\
	   &   & \times \left[ {1 \over \sqrt r} - {1 \over \sqrt{ r +
\Lambda^2 (1 + (f/2) \sin^2 2\varphi)} } \right] \nonumber \\
	& \approx & {u^2 \over 8\pi^2 \sqrt{2fr}}
\left[ \pi - 2\tan^{-1} {\sqrt r \over \Lambda} \right] \sim
	\left\{ \begin{array}{lr} 
	r^{-1/2}, & r < \Lambda^2 \\
	\Lambda/r, & r > \Lambda^2.
		\end{array} \right. \nonumber
\end{eqnarray}
In the second, approximate equality we have taken the large $f$ limit.
We see from this that the quasi one dimensional structure does not
affect the degree of divergence as $r \to 0$; indeed $f$ only affects
prefactors and not the scale $\Lambda^2$ to which $r$ should be compared.
To summarize, a mean-field treatment of the model \eq{action} should be
qualitatively correct except in the case of a $d=2$ $XY$ ferroelectric,
and we exclude this case henceforth.

\begin{figure}
\includegraphics[scale=0.47]{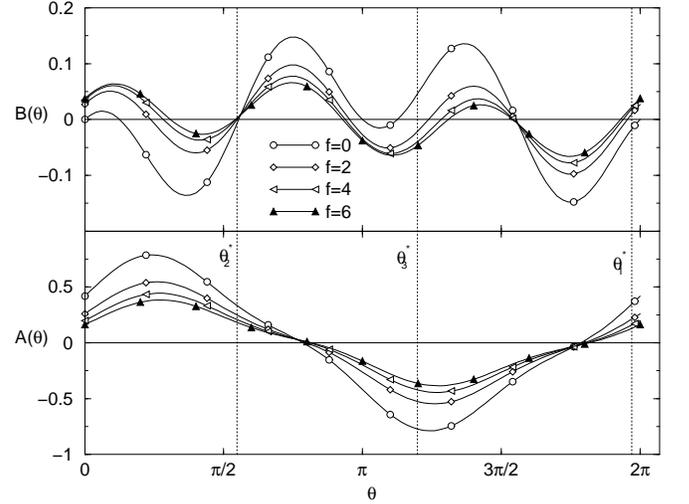}
\caption{Coefficients of the angular and radial part of the interaction
renormalization coefficient \eq{rthetarg} at $T=0$.  The stable roots of
$B(\theta)$ which determine the nature of the fixed points are marked by
vertical dotted lines.} \label{fixedpts}
\end{figure}

\begin{figure}
\includegraphics[scale=0.54]{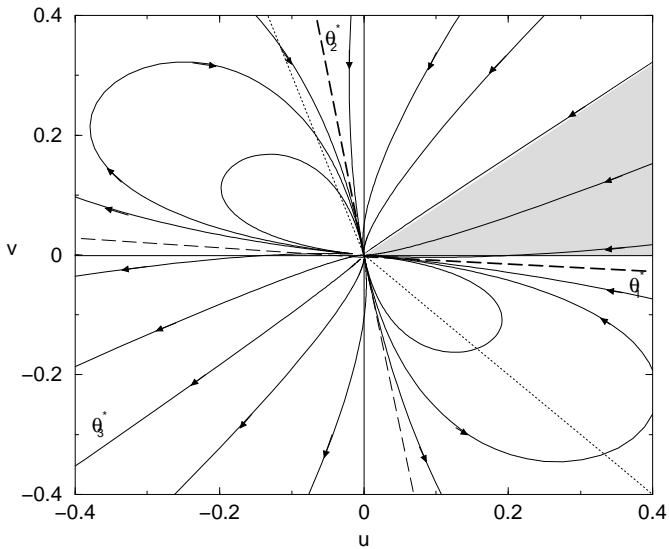}
\caption{Phase portrait for \eq{rthetarg} with  $f=1.1$.  The heavy dashed
lines indicate the $u/v$ ratios corresponding to stable fixed points of
\eq{rthetarg}; their continuations (light dashed lines) mark the boundaries
of the unstable region where $\rho$ flows to large values.  The dotted
lines mark the mean-field stability boundaries \eq{initconditions}.
The shaded region indicates initial conditions ($u_0, v_0$) which start
with Ising-like polarization orientation but eventually flow to the fixed
line $\theta^\star_1$ with nearly Heisenberg polarization orientation along
(111).} \label{phaseportrait}
\end{figure}

We further study the fixed points of \eq{rgint} in its full anisotropic form.
The Gaussian propagator \eq{propagator} in the strong dipole
interaction limit $g \to \infty$ is
\begin{multline} \label{propagatorg->infty}
G^0_{\alpha \beta} (\varpi, q) = {1 \over r + \varpi^2 + q^2 + f q_\alpha^2} \\
\times \left( \delta_{\alpha \beta} - { q_\alpha q_\beta \over Q
\denom{\beta}} \right)
\end{multline}
where $Q$ is defined in \eq{Q}.  With the use of cubic symmetry, the
possible combinations of $w_{\alpha\beta\gamma\delta}$ are reduced to
\begin{multline} \label{wclean}
w_{\alpha\beta\gamma\delta} = \left[ \left( A_1 - A_2
\right) \delta_{\alpha\gamma} + A_2 \right] \delta_{\alpha\beta}
\delta_{\gamma\delta} \\
+ A_3 \left( 1 - \delta_{\alpha\beta} \right)
\left( \delta_{\alpha\gamma} \delta_{\beta\delta} + \delta_{\alpha\delta}
\delta_{\beta\gamma} \right)
\end{multline}
The $f$- and $T$-dependent integrals $A_{1,2,3}$ are calculated in the
Appendix.  Substituting \eq{wclean} in \eq{rgint} yields coupled nonlinear
renormalization equations for $u$ and $v$, similar to those written by
Aharony and Fisher \cite{Aharony73} for the classical case (but note that
Aharony and Fisher expanded the coefficients $A_i$ about the limit of small
anisotropy whereas we retain their full $f$ dependence).  The stability of
the Gaussian fixed point $u=v=0$ is most transparently analysed using polar
coordinates in the $(u,v)$ plane: $u=\rho\cos\theta$ and $v=\rho\sin\theta$.
The renormalization equations for $u$ and $v$ then become
\begin{subequations} \label{rthetarg}
\begin{align}
{\dd \rho \over \dd \ln \Lambda} &= - A(\theta,T) \rho^2 \label{rhodot} \\
{\dd \theta \over \dd \ln \Lambda} &= - B(\theta,T) \rho \label{thetadot}
\end{align}
\end{subequations}
The $\theta$- and $T$-dependent coefficients $A$ and $B$ are given
in the Appendix (\eq{ABcoeffs}), and their $T=0$ limit is plotted
in Fig. (\ref{fixedpts}).  These coefficients are to be evaluated at the
running temperature $T(\Lambda)=T_\text{phys}e^{\ln\Lambda}$ and are derived
on the assumption that the physics is dominated by the (Gaussian) quantum
critical point; in other words, on the assumption that control parameter
$r$, interaction amplitude $\rho$ and temperature are not too large. In
particular, the model exhibits a pase transition at a temperature $T_c(r)$
discussed in detail below.  At temperatures sufficiently near to $T_c(r)$
a crossover to physics controlled by a classical, non-Gaussian critical point
will occur and the theory used here ceases to apply.  To estimate the region
of applicability of the equations presented here we follow \cite{Millis93},
noting first that the breakdown of the quantum critical theory will occur
in the classical region $T(\Lambda)>\Lambda$. In this regime the relevant
dimensionless interaction amplitude is $\rho_\text{classical}=T(\Lambda)
\rho(\Lambda)/\Lambda$ and \eq{rhodot} predicts the classical fixed
point $\rho^\star_\text{classical} = \lim_{T \to \infty} T / (\Lambda
A(\theta, T))$.  We find $\rho^\star_\text{classical} = \{1.25318,1.47604\}$
for the two fixed lines $\theta_1^\star$, $\theta_2^\star$ respectively,
shown in Fig. (\ref{phaseportrait}), and for $f=1.1$.  The $f$-dependence
of $\rho^\star_\text{classical}$ can be be summarized by the linear fits
$\rho^\star_\text{classicl} \approx 1.16 + 0.11 f$ for the $\theta_1^\star$
fixed line, and $\rho^\star_\text{classicl} \approx 0.48 + 0.77 f$ for
the $\theta_2^\star$ fixed line.  As an estimate of the range of validity
of the scaling equations, we argue they apply for $\rho_\text{classical}
\lesssim (1/2) \rho^\star_\text{classical}$, which corresponds to the $3D$
Ginzburg criterion $\rho_\text{classical}/r^{1/2}_\text{classical} \sim 1$.

We wish to study the stability of the fixed point $\rho=0$ in \eq{rthetarg}.
The solutions of \eq{rthetarg} asymptotically approach the fixed point along
`invariant lines' $\dot \theta = 0$ given by the stable roots $\theta^\star$
of $B(\theta, T)$ (\eq{B}).  Respectively, \eq{rthetarg} has a stable fixed
point $\rho \to 0$ if $A(\theta^\star, T) > 0$.  The functions $A(\theta)$
and $B(\theta)$ are shown in Fig. (\ref{fixedpts}) for several values of
the anisotropy $f$.  It is seen that there are three `invariant lines' of
which $\theta^\star_3$ (the middle in Fig. (\ref{fixedpts})) is unstable
($A(\theta^\star_3) < 0$).  The two stable solutions are $\theta^\star_1
\lesssim 0$ corresponding to a nearly Heisenberg fixed point $|u| >> v$;
$u<0$, and $\theta^\star_2 \gtrsim \pi/2$ corresponding to an Ising-like
fixed point $v >> |u|$, $u<0$.  The dependence of $\theta^\star_{1,2}$ on $f$
is weak and does not change the qualitative behavior.  The nature of the
`fixed line' solutions is most clearly seen in Fig. (\ref{phaseportrait})
which shows the phase portrait of \eq{rthetarg}.  The stable fixed lines
$\theta^\star_{1,2}$ are shown by heavy dashed lines.  Above and to the right
of the light dashed lines the flows are stable ($\rho \to 0$); below and to
the left, unstable ($\rho \to \infty$).  The region of stability found in
the RG analysis is wider than that found in the mean field approximation
(\eq{initconditions} shown in Fig. (\ref{phaseportrait}) as light dotted
lines).  The physical content of the two fixed lines $\theta^\star_{1,2}$
is different: $\theta^\star_1$ corresponds to a nearly isotropic system
with polarization along (111) but a relatively weak barrier against
polarization reorientation ($v/|u| \sim 0.15$, but weakly $f$-dependent),
whereas $\theta^\star_2$ corresponds to a strongly anisotropic system
with polarization along (100) and a barrier of relative order unity.
It is seen in Fig.  (\ref{phaseportrait}) that there exists a range of
initial conditions in the shaded wedge between the $v=0$ axis and the
separatrix in the first quadrant, which start with initial values $v_0 >
0$ favoring Ising symmetry but eventually flow to the $\theta^\star_1$
fixed line with $v<0$ and Heisenberg polarization symmetry.

\begin{figure}
\includegraphics[scale=0.64]{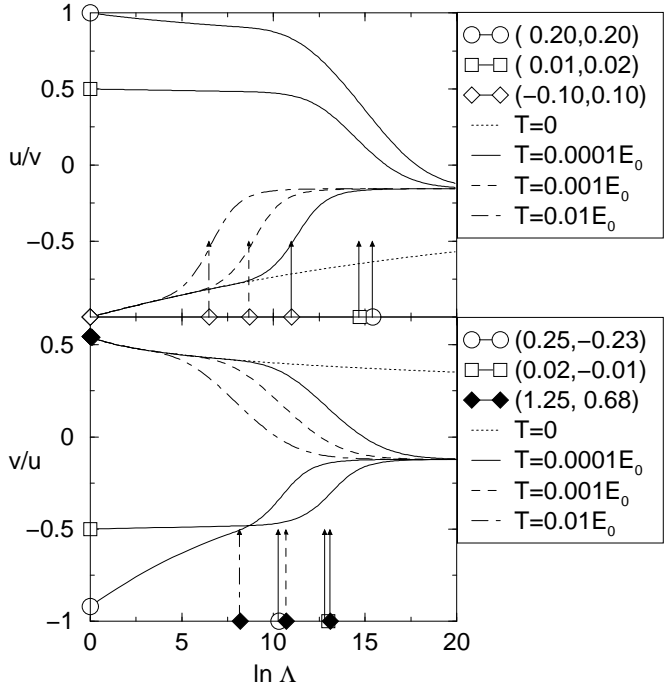}
\caption{The solutions of \eq{rthetarg} for $f=1.1$ plotted for several
different initial conditions.  Selected trajectories are shown for several
temperatures.  The trajectories converge into one of two `fixed lines'
also seen in Fig.  (\ref{phaseportrait}).  The top and bottom panels
show trajectories converging to $\theta_2^\star$ and $\theta_2^\star$
respectively.  The two panels are shown with inverted interaction constant
ratios relative to each other to capture the significant features of
the fixed line approach in each case.  The arrows mark the quantum --
classical crossover $\rho_\text{classical} \sim \rho^\star_\text{classical}$.
Trajectories that undergo polarization reorientation are marked with filled
symbols.} \label{ratios}
\end{figure}

We see that the ratio and even the sign of $u/v$ may change under
renormalization. In particular, for initial $u>0$, $v>0$ and $u/v$ less than
an ($f$-dependent) critical value of the order of unity, the {\em sign}
of $v$ changes under renormalization, corresponding to a predicted change
in the polarization direction as criticality is approached. Unfortunately,
the logarithmic nature of the scaling, combined with the numerically small
value of $B(\theta, T)$ and the factor of $\rho$ in \eq{thetadot} means that
at $T=0$ one must approach criticality extraordinarily closely to observe
the effect. The scaling turns out to be more rapid in the classical regime
$T(\Lambda)>\Lambda$, but as noted above our analysis cannot be extended
too far into this regime before the equations break down.

To further illustrate this point and to study how ``soon'' in RG
time this change of ordered state orientation occurs, and each fixed
line is reached, we show in Fig. (\ref{ratios}) the evolution of the
ratio of the two interaction constants along typical trajectories in
Fig. (\ref{phaseportrait}).  It is seen that the trajectories reach their
`fixed line' regime relatively late with long temperature dependent
transients sensitive to initial conditions.  Trajectories that start
with Ising symmetry $v_0 > 0$ and ultimately flow to a Heisenberg
fixed line with $v < 0$ are marked with filled symbols.  For example,
the initial conditions for BaTiO$_3$ are within the shaded range in Fig.
(\ref{phaseportrait}), and the sign reversal is expected to occur for $T =
0.0001 E_0$ at $\ln \Lambda \sim 15$.

The phase portrait Fig.  (\ref{phaseportrait})
changes only quantitatively with increased anisotropy, e.g. $f=5$: the
closed loop trajectories shrink towards the origin as higher anisotropy
reduces both interactions $u$ and $v$ (\eq{wsoverangles}), and the slope
of the fixed lines changes according to the $\theta$-roots shown in Fig.
(\ref{fixedpts}).

\section{Free energy, specific heat and mass renormalization}
\begin{figure}
\includegraphics[scale=0.47]{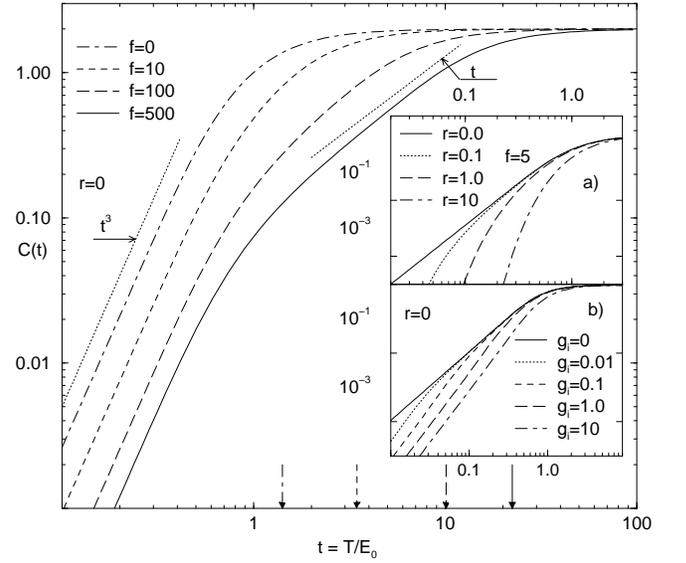}
\caption{Main figure: critical point specific heat per unit cell to
Gaussian order in units of $k_B$ for a $d=3$ Heisenberg model for a set
of anisotropy parameters $f$.  The vertical arrows mark the points $t =
\sqrt{2+f}$ for each respective curve.  Inset: a) the $r$-dependence
of $C(t)$ for $f=5$; b) Heisenberg to Ising crossover in specific heat,
for $f=0$.} \label{specHeatGraph}
\end{figure}

Within the Gaussian approximation the free energy per unit cell of the
system is given by
\begin{equation}
F = \sum_\lambda \int_q \left[ {1 \over 2} E_0
\varpi_\lambda (q) + T \ln \left( 1 - e^{- \varpi_\lambda (q) E_0/T}
\right) \right]
\end{equation}
where $\varpi_\lambda$ are the poles of \eq{propagator}.  The specific
heat can be obtained directly from this expression as
\begin{equation}
C =  -T {\partial^2 F \over \partial T^2} = \sum_\lambda \int {\dd^d q \over
(2\pi)^d} {\left[ \varpi_\lambda (q) E_0 / 2T \right]^2 \over \sinh^2 \left[
\varpi_\lambda (q) E_0 / 2T \right]}
\end{equation}
Using the general form of the eigenmodes in a cubic Heisenberg system with
anisotropy \eq{generalModes} and the isotropic Ising mode $\varpi^2(q) =
r + q^2 + g_I \cos^2 \theta$ the asymptotic low-temperature
behavior of the specific heat is
\begin{eqnarray}
C_H & = & \sum_{\lambda=1}^2 \int {\dd \Omega_q \over \pi^3} {(T/E_0)^3
\over \left[ 1 + f A_\lambda (\Omega_q) \right]^{3 / 2} } \int_0^\infty
{ x^2 \left( \kappa + x^2 \right) \dd x \over \sinh^2 \sqrt{\kappa + x^2}
} \nonumber \\
C_I & = & {3 \over \pi \sqrt{g_I}} \left( {T \over E_0} \right)^4
\int_0^\infty {t (\kappa + t) \over \sinh^2 \sqrt{\kappa + t}} \dd t
\label{asymptSpecHeat}
\end{eqnarray}
where $H$ and $I$ refer to Heisenberg and Ising respectively and $\kappa =
(E_0/2T)^2r$.  As is seen from \eq{asymptSpecHeat} in the low-temperature
limit the anisotropy $f$ enters only as a multiplicative factor in a
$f=0$ expression for the specific heat.  The specific heats of a $d=3$
Heisenberg model are shown in Fig. (\ref{specHeatGraph}) for $r=0$ (main
figure) and away from the critical point (inset).  We see that except in
the unrealistically strong ($f > 100$) case the only crossover visible is
from the quantal ($C \sim T^3$) to classical ($C \sim$ const) behavior as
$T$ is increased through the largest zone boundary phonon frequency (shown
by arrows in Fig. (\ref{specHeatGraph}).  The crossover from Heisenberg
to Ising symmetry is shown in inset b) of Fig.  (\ref{specHeatGraph})
for a set of Ising interaction strengths $g_I$.

\begin{figure}
\includegraphics[scale=0.47]{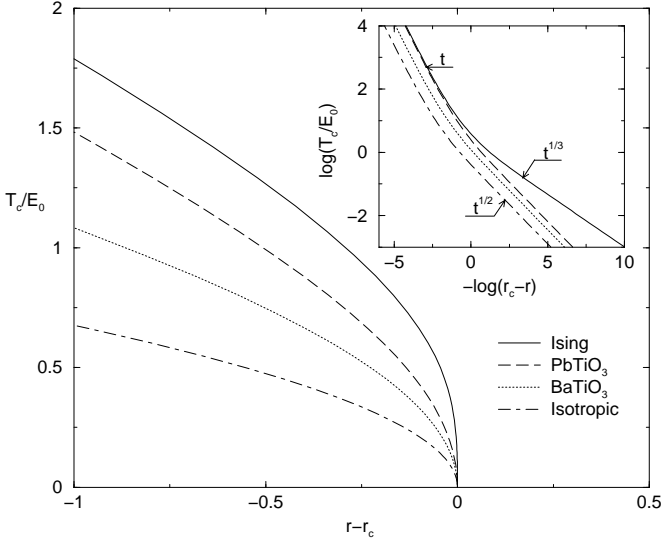}
\caption{Pressure dependence of the ferroelectric critical temperature
for the following cases: an Ising model with $g_I = 10$ and $v_0 = 1$;
PbTiO$_3$ and BaTiO$_3$ with parameters in Table \ref{parameters}; an
isotropic model with $f = 0$, $u_0 = 2$.}
\label{phaseDiagram}
\end{figure}

Finally we study the pressure dependence of the transition temperature.  The
mass flow equation is given by
\begin{equation} \label{massflow}
\dd r = 2 r(\Lambda) \ \dd \ln \Lambda + \dd R(\Lambda)
\end{equation}
where $\dd R(\Lambda)$ represents the one loop mass correction Fig.
(\ref{diagrams}b).  It is possible to express this diagram in terms of an
invariant of \eq{propagator}
\begin{equation} \label{massode}
\dd R(\Lambda) = \displaystyle{4 \over d} [(d+2)u(\Lambda) + 3v(\Lambda)]
T\sum_{\varpi_n}^\prime \int '{\dd^d q \over (2\pi)^d} \ \text{Tr} {\bf G^0}
\end{equation}
where the $q$-integration and Matsubara frequency summation are performed
in narrow shells of width $\dd \Lambda$ for each variable while the other
one is held fixed at the bandwidth cutoff, e.g. $\varpi_n \in [\Lambda,
\Lambda + \dd \Lambda]$ while $q = \Lambda$, and $q \in [\Lambda, \Lambda
+ \dd \Lambda]$ while $\varpi_n = \Lambda$.  The trace in \eq{massode}
is over the propagator eigenvalues \eq{eigenG}. Using the identity
$$
T\sum_{\varpi_n} {1 \over \varpi_n^2 + \omega_\lambda^2(q)} = \int_0^\infty
{\dd \omega \over 2 \omega} \coth {\omega \over 2T} \delta \left(\omega -
\omega_\lambda(q) \right)
$$
we obtain the solution of \eq{massflow} for $d=3$ in explicit form
\begin{align} \label{massRenorm}
r(\Lambda) =& e^{2\ln \Lambda} \left\{ r_0 + \int_0^{\ln \Lambda} \dd \ln
\Lambda' e^{-2\ln \Lambda'} 4\left[ {5 \over 3} u(\Lambda') + v(\Lambda')
\right] \right. \nonumber \\
	& \times {\Lambda^2 \over 16\pi^3} \sum_\lambda \int \dd \Omega
{1 \over \sqrt{1+fA_\lambda(\Omega)} } \\
	& \qquad \times \left. \left[ \coth {\Lambda
\sqrt{1+fA_\lambda(\Omega)} \over 2Te^{\ln \Lambda'}} + {\coth \Lambda /
2Te^{\ln \Lambda'} \over 1+fA_\lambda(\Omega)} \right] \right\} \nonumber
\end{align}
Here $r_0$ is the initial condition for the mass; $u(\Lambda)$ and
$v(\Lambda)$ are the solutions to \eq{rthetarg}; $A_\lambda(\Omega)$
are the angular dependent anisotropic factors of the dispersion from
\eq{generalModes}, and $T(\Lambda) = Te^{\ln \Lambda}$ is the flowing
temperature while $T$ is the real physical temperature.  We find $T_c$ from
the requirement that at the critical temperature the mass flows to zero:
$r(\Lambda \to \infty)$.  The quantum critical control parameter (which
in experimental realizations corresponds to e.g. hydrostatic pressure,
doping, etc.) is $r-r_c = r_0(T) - r_0(T=0)$.  We point out that since
a $3D$ ferroelectric is above its upper critical dimension we obtain
a qualitatively identical phase boundary if we simply use the initial
conditions $u_0$ and $v_0$ for the interaction constants in \eq{massRenorm}.
In that case $T_c(r-r_c)$ can be obtain from the simpler expression
\begin{equation} \label{massRenormSimple}
|r - r_c| = \sum_\lambda \int^\pi {d^d q \over (2\pi)^d} {{4 \over d} \left[
(d+2)u_0 + 3v_0 \right] \over \varpi_\lambda(q) \left[ \exp \left( {E_0
\over T_c} \varpi_\lambda \right) - 1 \right] }
\end{equation}
The phase boundary $T_c(r-r_c)$ obtained from \eq{massRenormSimple} is shown
in Fig.  (\ref{phaseDiagram}) for four representative sets of parameters.
All curves in Fig.  (\ref{phaseDiagram}) except Ising behave as $T_c \sim |r
- r_c|^{1/2}$ near $r_c$; the Ising behavior is $T_c \sim |r - r_c|^{1/3}$.
All curves cross over to $T_c \sim |r - r_c|$ as $T$ is increased through
the softest zone boundary phonon frequency $E_0$.

We briefly consider the effect of a small density of free carriers
characterized by an inter-carrier spacing $L_F$ and a diffusion constant
$D_F$. At length scales longer than $L_F$ and frequency scales lower than
$\omega_F = D_F/L_F^2$, these carriers will screen the interaction on the
scale $L_F$ and overdamp the dynamics. The details of the crossover depend
on the ratio $D_F/cL_F$. If $D_F/cL_F >> 1$, then there is a two-stage
crossover: as the scale is decreased, first the dynamics becomes overdamped
and then subsequently the characteristic length scale passes through the
screening length and the interaction becomes effectively short ranged. On the
other hand, if $D_F/cL_F << 1$, then screening and overdamping occur at the
same scale. Further studies of this crossover will be presented elsewhere.

All cases except for $2d$ $XY$ symmetry are above the upper critical
dimension enabling a controlled treatment.  Lattice-induced anisotropies
arising from the dipolar interaction are not small in real materials,
and lead e.g. to strong ``quasi one-dimensional'' effects in the phonon
spectrum (cf Fig. \ref{propGraph}).  However, we showed that for systems
above the upper critical dimension the effect on the critical behavior
is unimportant; only for unrealistically strong anisotropies $f > 100$
is an intermediate quasi one-dimensional regime visible in the specific
heat.  A change of polarization direction under scaling is suggested for
BaTiO$_3$ near $T_c(p)$ and for $p$ sufficiently close to the critical
pressure at which $T_c \to 0$ (although the scaling equations break down
at approximately the scale of the anisotropy change).  We have presented
exact results, in physical units, for the phase boundary and specific heat.
For PbTiO$_3$ and BaTiO$_3$ quantum critical effects are dominant for $T <
50$K if the materials are tuned by pressure to the quantum critical point.

Acknowledgements: We thank K. M. Rabe for helpful conversations and NSF DMR
00081075 and the University of Maryland -- Rutgers MRSEC for support.

\section{Appendix}
\setcounter{equation}{0}
\renewcommand{\theequation}{A\arabic{equation}}
We calculate the one-loop diagrams \eq{wsdef} by using a diagonal
representation for the Gaussian propagator achieved through the rotation
matrix ${\bf R}$
\begin{subequations} \label{rotation}
\begin{align}
G^0_{\alpha\beta} &= R_{\alpha\sigma} g_\sigma(\varpi_n,q) R^{-1}_{\sigma\beta}
\intertext{where}
g_\sigma(\varpi_n, q) &= {1 \over \varpi_n^2 + \omega_\sigma^2(q)} = {1 \over
\varpi_n^2 + r_\sigma + q^2(1+fA_\sigma(\theta,\varphi))} \label{eigenG}
\end{align}
\end{subequations}
The one-loop integrals then become
\begin{widetext}
\begin{align}
\begin{split}
w_{\alpha\beta\gamma\delta} &= T\sum_{\varpi_n} \int {\dd^d q \over
(2\pi)^d} \sum_{\mu\nu} R_{\alpha\mu} {1 \over \varpi_n^2 + \omega_\mu^2(q)}
R^{-1}_{\mu\beta} R_{\gamma\nu} {1 \over \varpi_n^2 + \omega_\nu^2(q)}
R^{-1}_{\nu\delta} \\
	&= i\int_0^\infty {\dd \omega \over 2\pi} \int {\dd^d q \over
(2\pi)^d} \sum_{\mu\nu} R_{\alpha\mu} R^{-1}_{\mu\beta} R_{\gamma\nu}
R^{-1}_{\nu\delta} \Im {1 \over \left[ \omega^2 + \omega_\mu^2(q) \right]
\left[ \omega^2 + \omega_\nu^2(q) \right]} \coth {\omega \over 2T} \\
\end{split}
\end{align}
We perform the integration over the magnitude of $q$ in $d=3$ and obtain the
remaining integrals over angles only, which we then calculate numerically
\begin{equation} \label{wsoverangles}
w_{\alpha\beta\gamma\delta} =
\begin{cases}
\begin{split}
{1 \over 16\pi^3} {\dd \Lambda \over \Lambda} \sum_{\mu\nu} &\int \dd
\Omega_q R_{\alpha\mu} R^{-1}_{\mu\beta} R_{\gamma\nu} R^{-1}_{\nu\delta}
{1 \over f \left[ A_\nu(\Omega_q) - A_\mu(\Omega_q) \right]} \\
	& \times \left( {\coth \displaystyle{\Lambda
\sqrt{1+fA_\mu} \over 2T} + \coth \displaystyle{\Lambda \over 2T} \over
\sqrt{1+fA_\mu}} - {\coth \displaystyle{\Lambda \sqrt{1+fA_\nu} \over 2T}
+ \coth \displaystyle{\Lambda \over 2T} \over \sqrt{1+fA_\nu}} \right)
\end{split} &
\text{for \ } A_\mu(\Omega_q) \neq A_\nu(\Omega_q) \\
\begin{split}
{1 \over 16\pi^3} {\dd \Lambda \over \Lambda} \sum_{\mu\nu} &\int
\dd \Omega_q R_{\alpha\mu} R^{-1}_{\mu\beta} R_{\gamma\nu}
R^{-1}_{\nu\delta} {1 \over 2 \left[ 1+fA_\mu(\Omega_q) \right]^{3/2}} \\
	& \times \left[ \coth \displaystyle{\Lambda
\sqrt{1+fA_\mu} \over 2T} \left( 1 + {\displaystyle {\Lambda \over T}
\sqrt{1+fA_\mu(\Omega_q)} \over \sinh \displaystyle {\Lambda \over T}
\sqrt{1+fA_\mu(\Omega_q)}} \right) + \coth \displaystyle{\Lambda \over
2T} \right]
\end{split} & \text{for \ } A_\mu(\Omega_q) = A_\nu(\Omega_q)
\end{cases}
\end{equation}
The isotropic case $f=0$ is described by the second expression in
\eq{wsoverangles}.  For numerical calculations \eq{wsoverangles} is more
conveniently written as
\begin{equation} \label{wsoveranglesnumerical}
w_{\alpha\beta\gamma\delta} =
\begin{cases}
\begin{split}
{\Lambda \over 16\pi^3} \dd \Lambda \sum_{\mu\nu} &\int \dd
\Omega_q R_{\alpha\mu} R_{\beta\mu} R_{\gamma\nu} R_{\delta\nu}
{g_\mu g_\nu \over \sqrt{g_\mu} + \sqrt{g_\nu}} \\
	& \times \left( {\sqrt{g_\mu} \coth \Lambda/( 2T\sqrt{g_\mu}) -
	\sqrt{g_\nu} \coth \Lambda/( 2T\sqrt{g_\nu}) \over \sqrt{g_\mu} -
	\sqrt{g_\nu}} + \coth {\Lambda \over 2T} \right)
\end{split} &
\text{for \ } g_\mu \neq g_\nu \\
\begin{split}
{\Lambda \over 16\pi^3} \dd \Lambda \sum_{\mu\nu} &\int \dd \Omega_q
R_{\alpha\mu} R_{\beta\mu} R_{\gamma\nu} R_{\delta\nu} \left(
{1 \over 2} g_\mu^{3/2} \right) \\
	& \times \left[ \coth \displaystyle{\Lambda
\over 2T\sqrt{g_\mu}} \left( 1 + {\Lambda/(T\sqrt{g_\mu}) \over \sinh
\Lambda/(T\sqrt{g_\mu})} \right) + \coth \displaystyle{\Lambda \over 2T}
\right]
\end{split} & \text{for \ } g_\mu = g_\nu
\end{cases}
\end{equation}
\end{widetext}
Here $R_{\alpha\mu}$ is a matrix whose columns are the $\mu$-th eigenvector
and $g_\mu = \Lambda^2 [1+fA_\mu(\Omega_q)]$ is the $\mu$-th eigenvalue of
\eq{propagatorg->infty} both evaluated at $\varpi = 0$; $r=0$; $q=\Lambda$,
and both having an implicit angular dependence.  At low temperatures $T <
\Lambda$ \eq{wsoveranglesnumerical} reduces to
\begin{equation} \label{wsnumerical}
w_{\alpha\beta\gamma\delta} = {\Lambda \over 8\pi^3} \dd \Lambda
\sum_{\mu\nu} \int \dd \Omega_q R_{\alpha\mu} R_{\beta\mu} R_{\gamma\nu}
R_{\delta\nu} {g_\mu g_\nu \over \sqrt{g_\mu} + \sqrt{g_\nu}}
\end{equation}

The cubic-symmetric integrals $A_{1,2,3}$ appearing in \eq{wclean} are
defined as
\begin{subequations}
\begin{align}
A_1(b) &= T\sum_{\varpi_n} \int {\dd^d q \over (2\pi)^d} \ \left( G^{zz}
\right)^2 = w_{zzzz} \\
A_2(b) &= T\sum_{\varpi_n} \int {\dd^d q \over (2\pi)^d} \ G^{xx} G^{yy}
= w_{xxyy} \\
A_3(b) &= T\sum_{\varpi_n} \int {\dd^d q \over (2\pi)^d} \ \left( G^{xy}
\right)^2 = w_{xyxy}
\end{align}
\end{subequations}
and are calculated from \eq{wsoveranglesnumerical} numerically.

The angle-dependence of the $\rho$ and $\theta$ derivatives in \eq{rthetarg}
is given by
\begin{subequations} \label{ABcoeffs}
\begin{align}
A(\theta,T) = & \ a_1 \cos^3\theta + c_2\sin^3\theta \\
	&+ \sin\theta\cos\theta \big[
(b_1 + a_2)\cos\theta + (c_1+b_2)\sin\theta \big] \nonumber \\
B(\theta,T) = & \ a_2 \cos^3\theta - c_1\sin^3\theta \label{B} \\
	&+ \sin\theta\cos\theta
\big[(b_2-a_1) \cos\theta + (c_2-b_1)\sin\theta \big] \nonumber
\end{align}
\end{subequations}
where
\begin{align*}
a_1 &= 4(3A_1 + 4A_2 + 14A_3) & a_2 &= 16(A_1 - A_2 - 2A_3) \\
b_1 &= 8(5A_1 + 16A_3) 		& b_2 &= 48(A_1 - A_3) \\
c_1 &= 36A_3 			& c_2 &= 36(A_1 - A_3)
\end{align*}

\end{document}